\documentclass[aps,prl,twocolumn,groupedaddress,preprintnumbers]{revtex4-1}

\usepackage{amsmath}
\usepackage{graphicx}
\usepackage[utf8]{inputenc}
\usepackage{braket}
\usepackage{isotope}

\usepackage{hyperref}
\hypersetup{colorlinks, linkcolor = [rgb]{0,0.0,0.75}, citecolor = [rgb]{0,0.0,0.75}, urlcolor = [rgb]{0,0.0,0.75}}

\newcommand{\threef}{\text{SU(3)$_{\rm f}$}}

\makeatletter
\g@addto@macro\bfseries{\boldmath}
\makeatother

\begin{document}

\preprint{MITP-21-009}
\preprint{CERN-TH-2021-024}

\title{Weakly bound $H$ dibaryon from SU(3)-flavor-symmetric QCD}

\author{Jeremy~R.~Green}
\email{green@maths.tcd.ie}
\altaffiliation[Present address: ]{School of Mathematics and Hamilton Mathematics Institute, Trinity College, Dublin 2, Ireland}
\affiliation{Theoretical Physics Department, CERN, 1211 Geneva 23, Switzerland}

\author{Andrew~D.~Hanlon}
\email{ahanlon@bnl.gov}
\affiliation{Helmholtz-Institut Mainz, Johannes Gutenberg-Universit\"at, 55099 Mainz, Germany}
\affiliation{GSI Helmholtzzentrum f\"ur Schwerionenforschung, 64291 Darmstadt, Germany}
\affiliation{Physics Department, Brookhaven National Laboratory, Upton, New York 11973, USA}

\author{Parikshit~M.~Junnarkar}
\email{parikshit@theorie.ikp.physik.tu-darmstadt.de}
\affiliation{Institut f\"ur Kernphysik, Technische Universit\"at
Darmstadt, Schlossgartenstra{\ss}e 2, 64289 Darmstadt, Germany}

\author{Hartmut~Wittig}
\email{hartmut.wittig@uni-mainz.de}
\affiliation{PRISMA Cluster of Excellence and Institut f\"ur Kernphysik, University of Mainz, Becher Weg 45, D-55099 Mainz, Germany}
\affiliation{Helmholtz-Institut Mainz, Johannes Gutenberg-Universit\"at, 55099 Mainz, Germany}
\affiliation{GSI Helmholtzzentrum f\"ur Schwerionenforschung, 64291 Darmstadt, Germany}

\date{\today}

\begin{abstract}
We present the first study of baryon-baryon interactions in the
continuum limit of lattice QCD, finding unexpectedly large lattice
artifacts. Specifically, we determine the binding energy of the $H$
dibaryon at a single quark-mass point. The calculation is
performed at six values of the lattice spacing $a$, using O($a$)-improved
Wilson fermions at the SU(3)-symmetric point with $m_\pi=m_K\approx 420$\,MeV.
Energy levels are extracted by applying a variational method to correlation matrices of
bilocal two-baryon interpolating operators computed using the distillation technique.
Our analysis employs L\"uscher's finite-volume quantization condition to determine
the scattering phase shifts from the spectrum and vice versa,
both above and below the two-baryon threshold.
We perform global fits to the lattice spectra using parametrizations
of the phase shift, supplemented by terms describing discretization
effects, then extrapolate the lattice spacing to zero. The phase shift
and the binding energy determined from it are found to be strongly
affected by lattice artifacts.
Our estimate of the binding energy in the continuum limit of three-flavor QCD is
$B_H^\threef=4.56\pm1.13_{\rm stat}\pm0.63_{\rm syst}$\,MeV.
\end{abstract}

%\keywords{}

\maketitle

The $H$ dibaryon is a scalar six-quark state with flavor content
$uuddss$, originally proposed in 1977 by Jaffe~\cite{Jaffe:1976yi}.
Despite years of effort, experimental searches
have not produced any hard evidence for its
existence~\cite{Takahashi:2001nm, Ahn:2013poa, Kim:2013vym}.
However, an upper bound on its binding energy has been
derived from the observed production
and decay pattern of a doubly strange $\isotope[6][\Lambda\Lambda]{He}$
hypernucleus~\cite{Takahashi:2001nm,Ahn:2013poa}.

Studying the properties of a potential $\Lambda$-$\Lambda$ bound state
will help our understanding of the hadronic ($\Lambda$-$\Lambda$)
interaction, which is relevant for the physics of double hypernuclei,
neutron-rich matter and neutron stars.
Recently, experimental data for two-particle correlations in p-p, p-Pb
and Au-Au collisions \cite{Adamczyk:2014vca, Acharya:2018gyz,
Acharya:2019yvb} have been analyzed to constrain the
$\Lambda$-$\Lambda$ interaction and provide model estimates for the
binding energy of the $H$ dibaryon.
In addition, a dedicated experiment is planned to search for it
at J-PARC~\cite{J-PARCE424572:2021mtr}.
Other approaches to study the $H$ dibaryon include chiral effective
field theory~\cite{Haidenbauer:2011ah, Haidenbauer:2015zqb, Li:2018tbt,
  Baru:2019ndr} and lattice QCD.

Lattice QCD studies of dibaryons and baryon-baryon scattering are very challenging
because of the signal-to-noise problem~\cite{Parisi:1983ae, Lepage:1989hd}
and the complexity of contractions. In response to
an inconsistency between results in the nucleon-nucleon
sector~\cite{Iritani:2017rlk, Wagman:2017tmp}, there has been a
recent focus on improved baryon-baryon spectroscopy
methods~\cite{Francis:2018qch, Horz:2020zvv, Amarasinghe:2021lqa}.
This work goes beyond that to achieve control over all systematic
effects for the $H$-dibaryon channel at one unphysical quark mass point.

\begin{figure}
  \centering
  \includegraphics{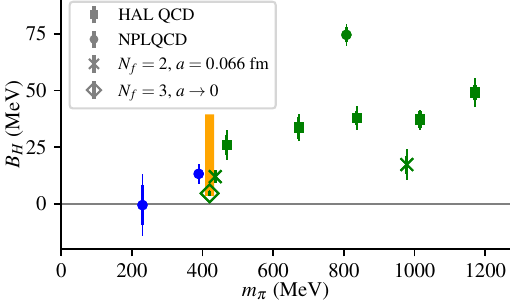}
  \caption{Binding energy versus pion mass: our results ---
    three-flavor QCD in the continuum [diamond,
    Eq.~\eqref{eq:bindE_final}] and two-flavor QCD at nonzero lattice
    spacing (crosses~\cite{supplement}) --- and published
    values~\cite{Beane:2010hg,Beane:2011zpa,Beane:2012vq,
      Inoue:2010es,Inoue:2011ai}. Green and blue symbols refer to
    SU(3)-symmetric and broken cases, respectively. The orange
    vertical band represents the range of binding energies obtained at
    nonzero lattice spacing for $N_f=3$.
    \label{fig:comparison}}
\end{figure}

There is a long history of calculations studying whether the $H$ dibaryon is a prediction of
QCD~\cite{Mackenzie:1985vv, Iwasaki:1987db, Pochinsky:1998zi,
  Wetzorke:1999rt, Wetzorke:2002mx, Luo:2007zzb, Luo:2011ar,
  Beane:2010hg, Beane:2011zpa, Beane:2011iw, Beane:2012vq,
  Inoue:2010hs, Inoue:2010es, Inoue:2011ai, Francis:2018qch,
  Sasaki:2019qnh}.
Results for the binding energy $B_H$ from these
calculations vary considerably, with estimates ranging from a few MeV
up to 75 MeV, depending on the methodology and/or the value of the
pion mass (see Fig.~\ref{fig:comparison}). Recently, employing near-physical pion
and kaon masses, the HAL~QCD Collaboration reported that the
$\Lambda$-$\Lambda$ interaction is only weakly attractive and does not
sustain a bound or resonant dihyperon \cite{Sasaki:2019qnh}.

In our previous work~\cite{Francis:2018qch}, using gauge fields
with dynamical $u$ and $d$ quarks and a quenched $s$ quark, we found
that the distillation method~\cite{Peardon:2009gh} produced a better
determination of the two-baryon spectrum than previously used methods.
At a heavy SU(3)-symmetric point with a pion mass of 960~MeV, we
obtained $B_H=19\pm 10$~MeV.

\begin{figure}
  \includegraphics{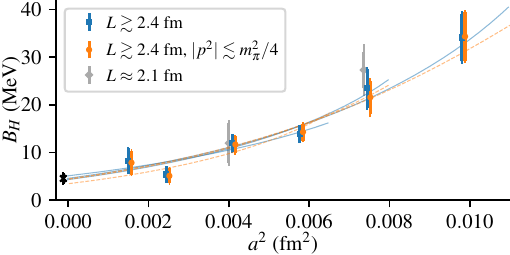}
  \caption{Binding energy versus squared lattice spacing, from fits to
    the full $p^2$ range (blue squares and solid curves) and to the
    near-threshold region (orange circles and dashed curves). Points
    are from fits to individual ensembles and curves are from the
    combined fits to the spectra of different subsets of the
    ensembles; they are not fitted to these points. Gray diamonds show
    results from the small-volume ensembles and the black
    cross shows our final estimate.}
  \label{fig:bindE_vs_a}
\end{figure}

In this letter we extend our calculations to lattice QCD with
dynamical $u$, $d$, and $s$ quarks with degenerate masses set to
their physical average value, corresponding to
$m_\pi=m_K\approx420$\,MeV \footnote{Preliminary results were presented
in~\cite{Hanlon:2018yfv}}. 
We present the first systematic study of discretization effects in a
multibaryon system, by computing finite-volume spectra at
several lattice spacings, extrapolating the
corresponding scattering phase shift to the continuum limit, and
determining the binding energy. As shown in Fig.~\ref{fig:bindE_vs_a},
at vanishing lattice spacing, we find
$B_H^\threef=4.56\pm1.30$~MeV, which is smaller than the result at the
coarsest lattice spacing by a factor of about 7.5. We
conclude that a thorough investigation of lattice artifacts is
indispensable for answering the question whether a bound $H$~dibaryon
exists in nature.

\begin{figure*}
  \centering
  \includegraphics{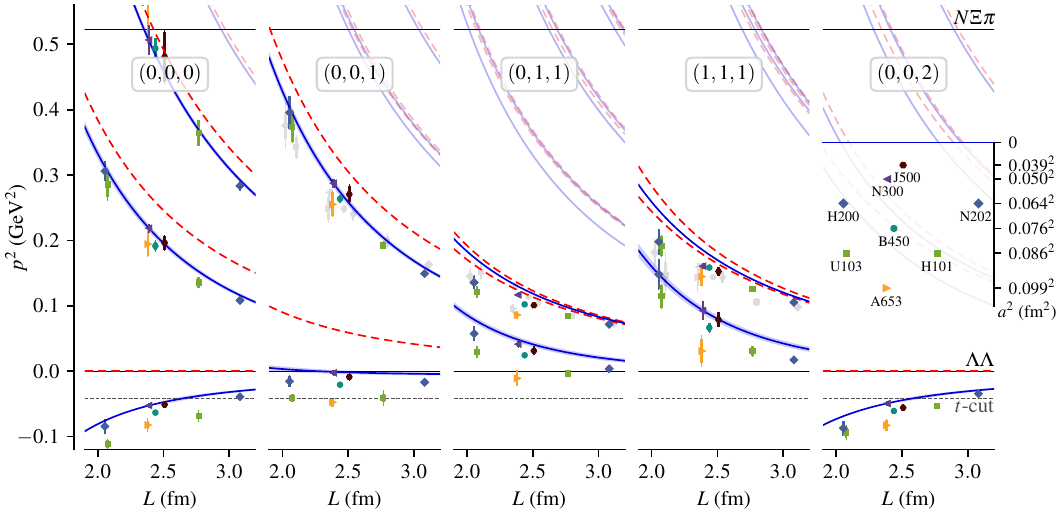}
  \caption{\label{fig:nf_3_spectra}Finite-volume spectrum:
    center-of-mass scattering momentum $p^2$ versus lattice extent
    $L$. The five different frames are shown separately and are
    labelled with $\vec D\equiv \vec P L/(2\pi)$.  Colored points show
    spin-zero levels and gray
    points (offset horizontally) show levels identified as spin one.
    Solid horizontal lines show
    the two- and three-particle thresholds while dashed horizontal
    lines represent the $t$-channel cut. The noninteracting spectrum
    is denoted by red dashed curves, and solid blue curves show the
    interacting spectrum determined in the continuum (see main text);
    the pale curves
    correspond to levels that have not been determined in the lattice
    calculation.
    The inset serves as a legend, showing $L$ and $a^2$ for the
    ensembles used in this work.}
\end{figure*}

Our calculations are based on a set of eight gauge ensembles
generated by CLS~\cite{Bruno:2014jqa}, with a
nonperturbatively O($a$)-improved Wilson-clover fermion action.
These ensembles have six different values of the
lattice spacing and multiple box sizes $L$ (all satisfying
$m_\pi L\geq 4.4$) as shown in
the inset of Fig.~\ref{fig:nf_3_spectra}~\cite{supplement}.

For each ensemble, we determine the energy levels in the rest frame
and in four moving frames. To this end, in each frame we compute a
Hermitian matrix of two-point correlation functions from a basis of
interpolating operators:
$C_{ij}(t) \equiv \langle \mathcal{O}_i(t) \mathcal{O}_j^\dagger(0) \rangle$.
The finite-volume spectrum $\{E_n\}$ determines the
exponential fall-off of $C_{ij}(t)$.

The building blocks of our operator basis are products of two
single-baryon operators projected to momenta $\vec p_1$ and $\vec p_2$
with total spin zero or one. For each frame momentum
$\vec P = \vec p_1 + \vec p_2$, we take linear combinations that
transform under the trivial irreducible representation of the little
group of $\vec P$, which contains the ${}^1S_0$ scattering channel~\cite{supplement}. Following
Refs.~\cite{Inoue:2010hs,deSwart:1963pdg}, the flavor content of our
interpolating operators is a linear combination of
isospin-zero $\Lambda\Lambda$, $\Sigma\Sigma$, and symmetric $N\Xi$
that corresponds to the singlet irreducible representation
of SU(3)-flavor.

Calculating the correlation functions of bilocal operators requires
the ability to compute ``timeslice-to-all" quark propagators. As in
our previous study \cite{Francis:2018qch}, we have used the
distillation technique~\cite{Peardon:2009gh,supplement}.

The finite-volume energy levels in each frame are determined by
solving a generalized eigenvalue problem (GEVP)~\cite{Luscher:1990ck, Blossier:2009kd, supplement},
$C(\tau_D) v_n = \lambda_n C(\tau_0) v_n$,
for fixed $\tau_0$ and $\tau_D$ satisfying $\tau_D > \tau_0\geq \tau_D/2$.
We then use the eigenvectors $v_n$ to construct
$\tilde C_{nm}(t)\equiv v_n^\dagger C(t) v_m$, an approximately
diagonalized correlator matrix.
We have verified that different combinations of
$(\tau_0,\tau_D)$ yield consistent results across a wide range of
values~\cite{supplement}.

Before fitting to the data, we divide the rotated two-baryon
correlators by a product of two single-baryon correlators that form
the corresponding two-baryon noninteracting level
$R_n (t) \equiv \tilde{C}_{nn} (t)/ [C_{\Lambda}^{\vec p_1} (t) C_{\Lambda}^{\vec p_2} (t)]$,
where $C_{\Lambda}^{\vec p_{i}}$ is a single-$\Lambda$ correlator
with momentum $\vec p_{i}$, and the total frame
momentum is $\vec p_1+\vec p_2$. The leading term in this ratio falls off
exponentially with the shift $\Delta E$ of the interacting
two-baryon energy away from the noninteracting level. In the ratio, we observe
a partial cancellation of correlated statistical fluctuations
and residual contributions from excited states,
which helps in the reliable determination of $\Delta E$.

Our finite-volume energies are determined from
single-exponential fits to $R_n(t)$.
For all levels, we choose $t_{\rm min}$, i.e.\ the smallest time
separation included in the fits, to lie in the plateau region of
$R_n(t)$. We also aim to have $t_{\rm min}$ lie in the plateau region
of the single-baryon correlators, and in the majority of cases we set
it to be the first time separation in this plateau region. Since the
single-baryon correlators take longer than the two-baryon correlators
to reach their asymptotic behavior, this ensures that all correlators
entering the ratio have little to no excited-state contamination.
In some cases, however, the signal of $R_n(t)$ is already
significantly degraded at the start of the single-baryon plateau
region, and we are led to choose a slightly lower $t_{\rm min}$ that
still lies within the plateau region of the correlator ratio.
For all levels, we estimate the sensitivity to $t_{\rm min}$ by
extracting an alternative spectrum with $t_{\rm min}$ further lowered,
and use it in subsequent analyses to estimate the systematic
uncertainty of our energy determination.

The fits also yield the couplings between each energy eigenstate and our operators.
For each frame that includes a spin-one operator,
we find one eigenstate that has strong overlap with only that operator,
allowing for a simple identification of the spin-one dominated states.

\begin{figure}
  \includegraphics[width=\columnwidth]{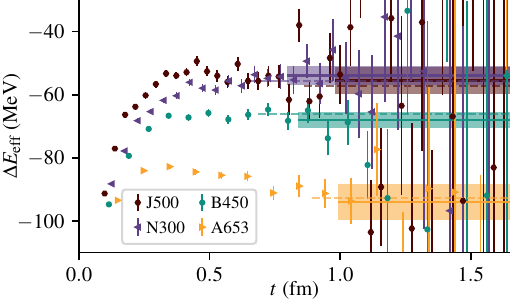}
  \caption{\label{fig:eff_energy_P1_E0}Effective energy difference
    obtained from $R_0(t)$ for the ground state in frame $(0,0,1)$
    on four ensembles with similar volumes. The bands
    show $\Delta E$ determined from a single-exponential fit to
    $R_0(t)$ and also indicate the range of $t$ used for the fit.
    The dashed lines show the alternative fit used to estimate
    systematic uncertainty.}
\end{figure}

Figure~\ref{fig:eff_energy_P1_E0} shows the effective energy
difference $\Delta E_{\rm eff}(t)\equiv -\frac{d}{dt}\log R(t)$
and the extracted $\Delta E$ for
the ground state in frame $(0,0,1)$ on four ensembles that differ
primarily in their lattice spacing. This level is particularly
important because it is the closest to the bound-state pole determined
in the phase shift analysis. An overview of the finite-volume spectrum
is shown in Fig.~\ref{fig:nf_3_spectra}, where the energy shifts are
transformed to the center-of-mass momentum $p$.
For every level, these
two figures show a clear increasing trend as the lattice spacing is reduced,
indicating that discretization effects are significant.

Given the two-particle scattering amplitude, Lüscher's finite-volume
quantization condition~\cite{Luscher:1990ux} and its
generalizations~\cite{Rummukainen:1995vs, Briceno:2013lba,
  Briceno:2014oea} determine the finite-volume spectrum, up to
exponentially suppressed corrections, between the
$t$-channel cut ($p^2 > -m_\pi^2/4$) and the three-particle threshold
($E_\text{cm} < 2m_B + m_\pi$). Since the quantization condition is
diagonal in spin~\cite{Briceno:2013lba, Briceno:2014oea}, the spin-one
part of the scattering amplitude does not affect the spin-zero
finite-volume spectrum, and we choose to ignore the spin-one states. In
addition, we neglect higher partial waves starting from ${}^1D_2$. In
this case, the quantization condition yields the ${}^1S_0$ phase shift
$\delta(p)$ at the momentum corresponding to each finite-volume energy
level:
\begin{equation}\label{eq:quantization}
  p\cot\delta(p) = \frac{2}{\sqrt{\pi}L\gamma}
  Z_{00}^{\vec PL/(2\pi)}\left(1, \left(\frac{pL}{2\pi}\right)^2 \right),
\end{equation}
where $\gamma=E/E_\text{cm}$ and $Z_{00}^{\vec D}$ is a generalized
zeta function. In addition to excluding levels with too-low or
too-high $p^2$ from our analysis, we must also exclude the first
excited levels in frames $(0,1,1)$ and $(1,1,1)$, as the ${}^1D_2$
partial wave is necessary to describe their position below the lowest
noninteracting level~\cite{supplement}.

The quantization conditions do not take discretization effects into
account; strictly speaking, they are only valid in the continuum.
There is no general formalism for finite-volume quantization at
nonzero lattice spacing, except for a simple model studied in
Ref.~\cite{Korber:2019cuq}. In principle, discretization effects would
affect both the scattering amplitude and the finite-volume
quantization condition. Effects on the former could include
$a$-dependence and frame-dependence of the scattering amplitude, as
well as couplings between $J^P$ that are forbidden in the continuum.
Effects on the latter could include a modification of the zeta
functions~\cite{Korber:2019cuq}. Either way, discretization effects
might spoil the factorization that separates spin-zero from spin-one.
Lacking a rigorous understanding, we have elected to model
discretization effects in a simple way, by allowing the parameters of
the phase shift to depend on $a$.

Our primary analysis is based on \emph{combined fits} of the
dependence of the phase shift on both $p^2$ and $a$. Specifically, our
model is
\begin{equation}\label{eq:combined_fits}
  p\cot\delta(p) = \sum_{i=0}^{N-1} c_i p^{2i},\quad
  c_i = c_{i0} + c_{i1} a^2.
\end{equation}
Concerning the dependence on $p^2$, we fit in two ways. The first
uses the near-threshold region, $|p^2|\lesssim m_\pi^2/4$ (where
the effective range expansion converges), with $N=2$
terms for the dependence on $p^2$. The second uses the full $p^2$
range, starting from the $t$-channel cut and stopping just below the
three-particle threshold, with $N=3$. Given $\{c_{ij}\}$, solving
Eq.~\eqref{eq:quantization} yields a discrete spectrum of $p^2$ for
each volume and frame; we fit these to the lattice spectra. For
comparison, we also performed fits to individual ensembles, neglecting
discretization effects. Given $\delta(p)$, a solution below threshold
to $p\cot\delta(p) = -\sqrt{-p^2}$ corresponds to a bound state pole.
All of the fits yielded a bound $H$ dibaryon.

Our preferred fit is to all ensembles using the full $p^2$ range; the
corresponding continuum interacting energy levels are shown as blue
curves in Fig.~\ref{fig:nf_3_spectra}. In addition to the alternative
spectrum fit range, we estimate the systematic uncertainty using the
root-mean-square difference of alternative combined fits that cover
all combinations of cuts on $p^2$ (full range or near threshold), $a$
(all six or the finest four), and $L$ (excluding $L\approx 2.1$~fm or
not). All of these fits have acceptable fit quality, with $p$-values
between 0.2 and 0.9. We explored adding an $a^3$ term in
Eq.~\eqref{eq:combined_fits} but found that this reduces $\chi^2$ by
at most 1.1 for each additional fit parameter, a sign of
overfitting. 

\begin{figure}
  \includegraphics{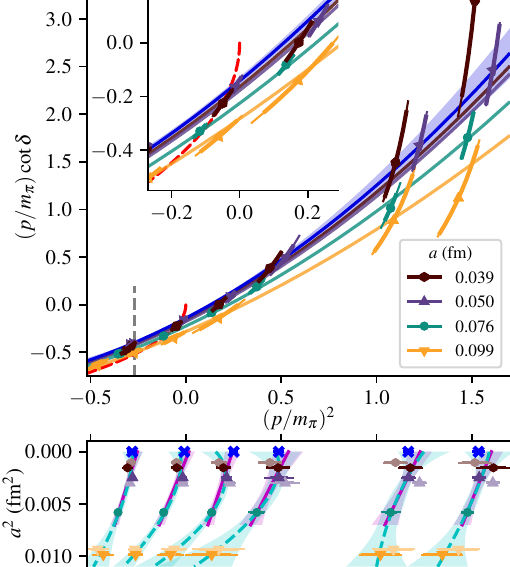}
  \caption{\label{fig:phase_shift}\textbf{Upper panel}: $p\cot\delta$
    versus $p^2$, normalized using the pion mass, with inset showing
    the near-threshold region. Data are shown for the four ensembles
    with $L\approx 2.4$~fm. Curves show the result from a combined
    fit, at nonzero lattice spacing (indicated by color)
    and in the continuum (blue with error band); intersections
    with the red dashed curve correspond to bound-state poles. Only
    points to the right of the vertical dashed line are included in
    the fit. \textbf{Lower panel}: level-by-level cross
    check of continuum extrapolation, with adjustments on three
    ensembles to match the target volume $L^*$.  Pale
    points (displaced vertically) show the levels before
    adjustment. The spectrum obtained from the continuum phase shift
    is indicated using blue crosses. Curves show
    continuum extrapolations of the form $b_0+b_1a^2$ excluding the
    coarsest lattice spacing (solid magenta) and $b_0+b_1a^2+b_2a^3$
    using all four lattice spacings (dashed cyan).}
\end{figure}

The phase shifts from the preferred fit, in the continuum and at
nonzero lattice spacing corresponding to the four ensembles with
$L\approx 2.4$~fm (J500, N300, B450, A653), are shown in
Fig.~\ref{fig:phase_shift}. Since these ensembles have similar values of
$L$, they allow us to perform a cross check, shown in the lower panel.
We select the volume of ensemble B450 as our target and call this box
size $L^*$. For the three other lattice spacings, we estimate each
energy level at $L^*$ by shifting from $L$ using the quantization
condition: $p^2(L^*) \approx p^2(L) + p^2_\text{q.c.}(L^*) -
p^2_\text{q.c.}(L)$. For each energy level, we then study the
dependence of $p^2(L^*)$ on the lattice spacing and compare it with
the value obtained from applying the quantization condition to the
continuum limit of the preferred fit.
The cross check shows that a level-by-level continuum extrapolation at
$L^*$ is consistent with the latter. However, some levels show
curvature in the dependence on $a^2$ and the fixed-$L^*$ extrapolation
is less precise, making it less useful than the combined fits.

Near threshold, we can write
$p\cot\delta=-1/a_0 + r_0p^2/2 + O(p^4)$, where $a_0$ is the
scattering length and $r_0$ is the effective range. We obtain
\begin{align}
  a_0^\threef &= 3.30 \pm 0.36 \pm 0.21 \text{ fm}, \label{eq:a0}\\
  r_0^\threef &= 0.98 \pm 0.04 \pm 0.05 \text{ fm},
\end{align}
where the first error is statistical and the second is systematic.
The dependence of the $H$ dibaryon binding energy on $a$ is shown in
Fig.~\ref{fig:bindE_vs_a}; in the continuum, we obtain
\begin{equation}\label{eq:bindE_final}
  B_H^\threef = 4.56 \pm 1.13 \pm 0.63 \text{ MeV},
\end{equation}
which is substantially lower than the binding energies determined at
nonzero lattice spacing, except on the finest two of our ensembles.

We have reported the first lattice study of a baryon-baryon system in the
continuum limit. The crucial elements of our methodology are the
finite-volume quantization condition, supplemented by terms describing
discretization effects and applied over a wide range of lattice
spacings, as well as the subsequent extrapolation to the continuum
limit. We conclude that cutoff effects are large and cannot be
ignored in an investigation of the $H$ dibaryon using lattice QCD; it
will be essential to study their importance in other multibaryon
systems such as the deuteron, where calculations
disagree~\cite{Iritani:2017rlk, Wagman:2017tmp, Horz:2020zvv}.
Our final result for the binding energy, given in
Eq.~\eqref{eq:bindE_final}, suggests the existence of a weakly bound
$H$~dibaryon, which is not only at variance with Jaffe's original bag
model prediction \cite{Jaffe:1976yi} of a deeply bound $uuddss$
state, but is also substantially lower than the binding energies
determined in previous lattice calculations \cite{Beane:2010hg,
Beane:2011zpa, Beane:2011iw, Beane:2012vq, Inoue:2010hs, Inoue:2010es,
Inoue:2011ai, Francis:2018qch} at nonzero lattice spacing
(see Fig.~\ref{fig:comparison}).
This adds to the evidence against deeply bound hexaquark dark
matter~\cite{Farrar:2017eqq, Gross:2018ivp, Farrar:2018hac,
  Kolb:2018bxv, McDermott:2018ofd, Echenard:2018rfe, Azizi:2019xla,
  Farrar:2020zeo}.
  An obvious caveat is that our calculation was performed
  for one set of degenerate quark masses. The issue of SU(3) symmetry
  breaking --- which is crucial, since the splitting between physical
  $\Lambda\Lambda$ and $N\Xi$ thresholds is larger than
  $B_H^\threef$ --- is currently under
  investigation~\cite{Padmanath_Lat21}. Previous estimates based on
  extrapolations of lattice data found a
  bound state at the physical point unlikely~\cite{Shanahan:2011su,
    Haidenbauer:2011ah, Inoue:2011ai,
    Yamaguchi:2016kxa, Li:2018tbt}; our smaller binding energy should
  make it even less likely.

\begin{acknowledgments}
  We thank Maxwell T.\ Hansen, Ben H\"orz, and Daniel Mohler for many helpful conversations.
  Calculations for this project used resources on the supercomputers
  JUQUEEN~\cite{juqueen}, JURECA~\cite{jureca}, and JUWELS~\cite{juwels} at Jülich
  Supercomputing Centre (JSC). The authors gratefully acknowledge the
  support of the John von Neumann Institute for Computing and Gauss
  Centre for Supercomputing e.V.\ (\url{http://www.gauss-centre.eu})
  for project HMZ21. 
  The raw distillation data were computed using
  QDP++~\cite{Edwards:2004sx}, PRIMME~\cite{PRIMME}, and the deflated
  SAP+GCR solver from openQCD~\cite{openQCD}. Contractions were
  performed with a high-performance BLAS library using the Python
  package opt\_einsum~\cite{opt_einsum}.
  The correlator analysis was done using SigMonD~\cite{sigmond}.
  Much of the data handling and the subsequent phase shift analysis was
  done using NumPy~\cite{numpy} and SciPy~\cite{scipy}. The
  plots were prepared using Matplotlib~\cite{Hunter:2007}.
  This research was partly supported by Deutsche
  Forschungsgemeinschaft (DFG, German Research Foundation) through the
  Cluster of Excellence ``Precision Physics, Fundamental Interactions
  and Structure of Matter'' (PRISMA+ EXC 2118/1) funded by the DFG
  within the German Excellence Strategy (Project ID 39083149), as well
  as the Collaborative Research Centers SFB 1044 ``The low-energy frontier
  of the Standard Model'' and
  CRC-TR 211 ``Strong-interaction matter under extreme conditions'' (Project
  ID 315477589 -- TRR 211). 
  ADH is supported by: (i) The U.S. Department of Energy, Office of Science,
  Office of Nuclear Physics through the Contract No. DE-SC0012704 (S.M.);
  (ii) The U.S. Department of Energy, Office of Science, Office of Nuclear
  Physics and Office of Advanced Scientific Computing Research, within the
  framework of Scientific Discovery through Advance Computing (SciDAC)
  award Computing the Properties of Matter with Leadership Computing Resources.
  We are grateful to our colleagues
  within the CLS initiative for sharing ensembles.
\end{acknowledgments}

\bibliographystyle{apsrev4-1}
\nocite{revcontrol}
\bibliography{Hdibaryon_3f.bib}

% put \end{document} here to exclude the supplemental material

% balance columns at end of main text
\onecolumngrid
\clearpage
\twocolumngrid

\begin{center}
  {\large \bf Supplemental material}
\end{center}

% turn on section numbers for the supplement, when it's included in the document
\makeatletter
\c@secnumdepth=4
\makeatother

\newif\ifsepsupp
\sepsuppfalse

\setcounter{equation}{0}
\setcounter{figure}{0}
\setcounter{table}{0}
\renewcommand{\theequation}{S\arabic{equation}}
\renewcommand{\thefigure}{S\arabic{figure}}
\renewcommand{\thetable}{S\Roman{table}}
% also need the following so that hyperref links aren't broken
\renewcommand{\theHequation}{S\arabic{equation}}
\renewcommand{\theHfigure}{S\arabic{figure}}
\renewcommand{\theHtable}{S\Roman{table}}

\ifsepsupp
\newcommand{\refeqquantization}{(1)}
\newcommand{\refeqbindEfinal}{(5)}
\newcommand{\reffigspectra}{3}
\newcommand{\reffigbindEvsa}{1}
\newcommand{\reffigeffenergy}{2}
\newcommand{\reffigcomparison}{5}
\else
\newcommand{\refeqquantization}{\eqref{eq:quantization}}
\newcommand{\refeqbindEfinal}{\eqref{eq:bindE_final}}
\newcommand{\reffigspectra}{\ref{fig:nf_3_spectra}}
\newcommand{\reffigbindEvsa}{\ref{fig:bindE_vs_a}}
\newcommand{\reffigeffenergy}{\ref{fig:eff_energy_P1_E0}}
\newcommand{\reffigcomparison}{\ref{fig:comparison}}
\fi

In this supplement, we provide additional details for our
calculation. Section~\ref{app:details} specifies the lattice action
and ensembles. The precise definitions of our interpolating operators
are given in Section~\ref{app:operators}. Our implementation of the
distillation approach is described in
Section~\ref{app:distillation}. We provide further details about our
determination of the spectrum in Section~\ref{app:corr} and our
combined fits to the spectra at different lattice spacings in
Section~\ref{app:combined}. The analysis of two $N_f=2$ ensembles is
provided in Section~\ref{app:Nf2}. Finally, Section~\ref{sec:data}
describes the spectrum data being made available with this article.

\section{Lattice ensembles\label{app:details}}

Our calculations are based on a set of gauge ensembles with $N_f=2+1$
flavors of dynamical quarks, generated by CLS using the openQCD code
suite \cite{Luscher:2012av} and listed in Table~\ref{tab:ensembles}.
The fields are described by the
tree-level O($a^2$)-improved L\"uscher-Weisz action and the O($a$)-improved
Wilson-Clover action in the quark sector, with the
improvement coefficient $c_{\rm sw}$ tuned to the nonperturbative
determination of Ref.\,\cite{Bulava:2013cta}. Open or periodic
boundary conditions in the time direction are employed. All ensembles
realize SU(3) symmetry, with $m_\pi=m_K\approx 420$\,MeV, at
six different values of the lattice spacing, covering a range
between 0.04 and 0.1\,fm. Here we also take the opportunity to extend
our earlier calculations with $N_f=2$ flavors of dynamical
quarks~\cite{Francis:2018qch}. The respective simulation parameters
are listed in Table~\ref{tab:ensembles}, and a detailed description
can be found in Section~\ref{app:Nf2}.

As discussed in Ref.~\cite{Bruno:2016plf}, the quark masses are not
exactly matched among the different lattice spacings. Given our choice
of scale setting, this corresponds to a 3\% variation in the pion
mass, from 411 to 424~MeV. This is expected to produce a shift in the
octet baryon mass of order 10~MeV, preventing a simple study of
discretization effects in the octet baryon mass. However, the latter
also varies by just 3\% among our ensembles, which puts a likely upper
bound on the size of discretization effects. For our main study of
baryon-baryon interactions, we always determine energy differences
from noninteracting levels and convert them to $p^2$ using the baryon
mass determined on the same ensemble, cancelling the leading effect
due to slightly varying baryon masses. Our expectation is that the
mistuning of the pion mass will affect the energy differences at the
few-percent level, which is much smaller than our statistical
uncertainty.

\begin{table*}
  \caption{\label{tab:ensembles}Overview of lattice ensembles. Each
    ensemble is characterized by the gauge coupling parameter $\beta$,
    the quark hopping parameter $\kappa$, the lattice size, and the
    temporal boundary condition. For $N_f=3$, the lattice spacing $a$ was
    determined for the second-finest lattice spacing from the result
    in Ref.~\cite{Bruno:2016plf} and scaled to the other lattice
    spacings using the gradient flow scale $t_0$~\cite{Luscher:2010iy}
    determined at the symmetric point. For the ensembles with $N_f=2$
    we use the lattice spacing determined in Ref. \cite{Fritzsch:2012wq}.
    The masses of the light octets of pseudoscalar mesons and spin-1/2
    baryons are given by $m_\pi$ and $m_B$, respectively. On each of the
    $N_\text{conf}$ gauge configurations analyzed, $N_\text{tsrc}$ source
    timeslices were used. Including both forward and backward-propagating
    states, the total number of measurements used is
    $N_\text{meas}=N_tN_\text{conf}$, where $N_t=2(N_\text{tsrc}-N_\text{skip})$.
    To avoid boundary effects, we omit some potential measurements, such
    as the backward-propagating states from the first source timeslice; thus,
    $N_\text{skip}$ is 0 for the ensembles with periodic boundary
    conditions and between 1 and $N_\text{tsrc}/2$ for the ensembles
    with open boundary conditions. Finally, $N_\text{LapH}$ is the
    number of low modes of the Laplacian used in the
    Laplacian-Heaviside smearing.}
  \begin{ruledtabular}
  \begin{tabular}{l|ccllc|lccrc|rcrc}
    Label & $N_f$ & $\beta$ & $\kappa$ & size & bdy.\ cond.
    & $a$ (fm) & $m_\pi$ (MeV) & $L$ (fm) & $m_\pi L$ & $m_B$ (GeV)
    & $N_\text{conf}$ & $N_\text{tsrc}$ & $N_\text{meas}$ & $N_\text{LapH}$ \\\hline
    J500 & 3 & 3.85 & 0.136852 & $64^3\times 192$ & open
    & 0.0392 & 411 & 2.5 & 5.2 & 1.18
    & 1341 &12 & 24138 & 36\\
    N300 & 3 & 3.70 & 0.137 & $48^3\times 128$ & open
    & 0.0498 & 422 & 2.4 & 5.1 & 1.20
    & 2047 &12 & 24564 & 32\\
    N202 & 3 & 3.55 & 0.137 & $48^3\times 128$ & open
    & 0.0642 & 412 & 3.1 & 6.4 & 1.17
    &  899 & 8 & 10788 & 68\\
    H200 & 3 & 3.55 & 0.137 & $32^3\times 96$ & open
    & 0.0642 & 419 & 2.1 & 4.4 & 1.20
    & 2000 & 8 & 16000 & 20\\
    B450 & 3 & 3.46 & 0.13689 & $32^3\times 64$ & periodic
    & 0.0762 & 417 & 2.4 & 5.2 & 1.18
    & 1612 & 8 & 25762 & 32\\
    H101 & 3 & 3.40 & 0.13675962 & $32^3\times 96$ & open
    & 0.0865 & 417 & 2.8 & 5.9 & 1.16
    & 2016 & 4 & 12096 & 48\\
    U103 & 3 & 3.40 & 0.13675962 & $24^3\times 128$ & open
    & 0.0865 & 414 & 2.1 & 4.4 & 1.18
    & 5658 & 5 & 45264 & 20\\
    A653 & 3 & 3.34 & 0.1365716 & $24^3\times 48$ & periodic
    & 0.0992 & 424 & 2.4 & 5.1 & 1.17
    & 5050 & 4 & 40400 & 32\\
    \hline
    E5 & 2 & 5.30 & 0.13625& $32^3\times 64$ & periodic
    & 0.0658 & 437 & 2.1 & 4.7 & 1.29
    & 2000 & 4 & 16000 & 30\\
    E1 & 2 & 5.30 & 0.1355 & $32^3\times 64$ & periodic
    & 0.0658 & 979 & 2.1 & 10.4 & 2.03
    & 168 & 8 & 2688 & 30
  \end{tabular}
  \end{ruledtabular}
\end{table*}

\section{Interpolating operators\label{app:operators}}

In our previous study~\cite{Francis:2018qch}, we found that bilocal
two-baryon operators are more effective than local hexaquark operators
at identifying the low-lying spectrum; therefore, in this work we use
only the former. To begin, we define the single-octet-baryon
operators, which make use of the three-quark combination
\begin{equation}
  \label{eq:singlebaryon}
  [rst]_\alpha = \frac{1}{\sqrt{18}}\epsilon^{ijk}(s^T_i C\gamma_5 P_+ t_j)r_{k\alpha}.
\end{equation}
Here $r$, $s$, and $t$ denote smeared quark fields of generic flavor
at the same point and $P_+=(1+\gamma_0)/2$ is a positive-parity
projector. This satisfies $[rst]=-[rts]$ and
$P_+([rst]+[str]+[trs])=0$. The members of the the SU(3)-flavor
octet are defined following Ref.~\cite{Inoue:2010hs}:
\begin{equation}
\label{eq:octet}
\begin{gathered}
  n = [dud],\quad
  p = [uud],\\
  \Sigma^- = -[dds],\quad
  \Sigma^0 = \tfrac{-1}{\sqrt{2}}([dus]+[uds]),\quad
  \Sigma^+ = -[uus],\\
  \Lambda = \tfrac{1}{\sqrt{6}}(2[sud]-[uds]-[dsu]),\\
  \Xi^- = [ssd],\quad
  \Xi^0 = [ssu].
\end{gathered}
\end{equation}

The spin-zero and spin-one two-baryon operators are defined as follows:
\begin{align}
  [B_1B_2]_0(\vec p_1,\vec p_2) &= \sum_{\vec x,\vec y}
       e^{-i\vec p_1\cdot\vec x} e^{-i\vec p_2\cdot\vec y}
       B_1^T(\vec x)C\gamma_5P_+B_2(\vec y), \label{eq:spin0}\\
  [B_1B_2]_i(\vec p_1,\vec p_2) &= \sum_{\vec x,\vec y}
       e^{-i\vec p_1\cdot\vec x} e^{-i\vec p_2\cdot\vec y}
       B_1^T(\vec x)C\gamma_iP_+B_2(\vec y). \label{eq:spin1}
\end{align}
In these operators, the baryon $B_j$ is projected to momentum
$\vec p_j$, and the total momentum is $\vec P = \vec p_1+\vec
p_2$. Each operator constructed in this way can be identified with a
noninteracting finite-volume energy level of energy
$E=\sum_j\sqrt{m_{B_j}^2+\vec p_j^2}$. These operators satisfy the
exchange symmetry relations
\begin{align}
  [B_1B_2]_0(\vec p_1,\vec p_2)&=[B_2B_1]_0(\vec p_2,\vec p_1),\\
  [B_1B_2]_i(\vec p_1,\vec p_2)&=-[B_2B_1]_i(\vec p_2,\vec p_1).
\end{align}
This work is focused on flavor-symmetric channels, which implies that
the spin-zero operators are even under exchange of momenta and are
thus associated with even partial waves, and the opposite is true for
the spin-one operators. For each total momentum $\vec P$, we construct
operators that transform under the trivial ($A_1^+$ or $A_1$)
irreducible representation of the little group of $\vec P$, which
contains the ${}^1S_0$ scattering channel. Generically, these have the
form
\begin{align}
\text{(spin zero)}&\quad \sum_{j}   c_{j}  [B_1B_2]_0(\vec p_{j},\vec P-\vec p_{j}),\\
\text{(spin one)} &\quad \sum_{i,j} c_{ij} [B_1B_2]_i(\vec p_{j},\vec P-\vec p_{j}),
\end{align}
for some coefficients $c_j$ or $c_{ij}$. For each operator, we choose
$\{\vec p_j\}$ such that they lie in the group orbit of a reference
momentum $\vec p$ under the little group of $\vec P$. Representative
momenta $\vec p_1$ and $\vec p_2$ for each of our operators are listed
in Table~\ref{tab:operators}, and these operators are given explicitly
in the following subsections. In each frame, we make use of one operator for each
noninteracting level below a certain threshold. In the noninteracting
and nonrelativistic limit, in all cases the energy gap to the first
uncontrolled state, i.e.\ from the
highest level for which an operator is included to the lowest level
for which an operator is not included, is $(2\pi/L)^2/m_B$, except in
frame $\vec P=(2\pi/L)(1,1,1)$, where this gap is doubled.

\begin{table}
  \caption{\label{tab:operators}Two-baryon interpolating operators
    used in each frame. Each operator is indicated by the total spin
    and a representative combination of individual baryon momenta,
    $\vec p_1+\vec p_2$, given in units of $2\pi/L$.}
  \begin{ruledtabular}
  \begin{tabular}{l|l|l}
    Frame & Spin zero & Spin one \\\hline\hline
    $(0,0,0)$ $A_1^+$
          & $(0,0,0)+(0,0,0)$ & \\
          & $(0,0,1)+(0,0,-1)$ & \\
          & $(0,1,1)+(0,-1,-1)$ & \\ \hline
    $(0,0,1)$ $A_1$
          & $(0,0,1)+(0,0,0)$ & $(0,1,1)+(0,-1,0)$ \\
          & $(0,1,1)+(0,-1,0)$ & \\ \hline
    $(0,1,1)$ $A_1$
          & $(0,1,1)+(0,0,0)$ & $(0,0,1)+(0,1,0)$ \\
          & $(0,0,1)+(0,1,0)$ & \\ \hline
    $(1,1,1)$ $A_1$
          & $(1,1,1)+(0,0,0)$ & $(0,1,1)+(1,0,0)$ \\
          & $(0,1,1)+(1,0,0)$ & \\ \hline
    $(0,0,2)$ $A_1$ & $(0,0,1)+(0,0,1)$
  \end{tabular}
  \end{ruledtabular}
\end{table}

The flavor content of our chosen operators belongs to the strangeness
$-2$, isospin zero sector:
\begin{align}
  [\Lambda\Lambda]^{I=0} &= [\Lambda\Lambda],\\
  [\Sigma\Sigma]^{I=0} &= \frac{1}{\sqrt{3}}\left( [\Sigma^+\Sigma^-] - [\Sigma^0\Sigma^0] + [\Sigma^-\Sigma^+] \right),\\
  [N\Xi_s]^{I=0} &= \frac{1}{2}\left( [p\Xi^-] - [n\Xi^0] + [\Xi^-p] - [\Xi^0n] \right).
\label{eq:Qnumbers}
\end{align}
These are transformed to the singlet irreducible representation of flavor SU(3)
following Refs.~\cite{Inoue:2010hs,deSwart:1963pdg}:
\begin{equation}
\label{eq:singlet}
  [\textbf{1}] = -\sqrt{\frac{1}{8}}[\Lambda\Lambda]^{I=0}
                 +\sqrt{\frac{3}{8}}[\Sigma\Sigma]^{I=0}
                 +\sqrt{\frac{4}{8}}[N\Xi_s]^{I=0}.\\
\end{equation}

In the following subsections we list the spin-zero and spin-one flavor-symmetric
interpolators in the trivial irrep in each frame. Each moving frame
has several equivalent copies, related by lattice rotations; the
listed operators will be given in a generic way for all equivalent
frames, such that all operators in each irrep transform in the same
way between equivalent frames. (We have performed a cross-check using
computer algebra to verify these transformation properties.)
For each term $[B_1B_2](\vec p_1,\vec
p_2)$, only $\vec p_1$ will be given, since $\vec p_2=\vec P-\vec
p_1$. The operators will be labeled $[BB]_{\Lambda, \vec P
  L/(2\pi)}^{s(n_1,n_2)}$, where $\Lambda$ is the irrep, $s$ is the
spin, and $p_i^2=n_i(2\pi/L)^2$.

\subsection{(0,0,0) $A_1^+$}

Here we make use of the standard basis vectors $\vec e_i$.
\begin{align}
  [BB]_{A_1^+(0,0,0)}^{0(0,0)} &=  [BB]_0(\vec 0),\label{eq:zzz}  \\
  [BB]_{A_1^+(0,0,0)}^{0(1,1)} &= \frac{1}{\sqrt{3}}\sum_i [BB]_0(\tfrac{2\pi}{L}\vec e_i), \\
  [BB]_{A_1^+(0,0,0)}^{0(2,2)} &= \frac{1}{\sqrt{6}}\sum_i \sum_{j>i}
    \sum_{r\in\{\pm 1\}} [BB]_0(\tfrac{2\pi}{L}[\vec e_i+r\vec e_j]).
\end{align}

\subsection{(0,0,1) $A_1$}

Here the frame momentum is $\vec P=\pm \frac{2\pi}{L} \vec e_k$ for some $k$.
\begin{align}
  [BB]_{A_1(0,0,1)}^{0(0,1)} &= [BB]_0(\vec 0), \\
  [BB]_{A_1(0,0,1)}^{0(1,2)} &= \frac{1}{2}\sum_{i\neq k}\bigl(
    [BB]_0(\tfrac{2\pi}{L}\vec e_i)
 +  [BB]_0(-\tfrac{2\pi}{L}\vec e_i) \bigr), \\
  [BB]_{A_1(0,0,1)}^{1(1,2)} &= \frac{1}{2}\sum_{i\neq k}\bigl(
    [BB]_i(\vec e_i\times\vec P)
 -  [BB]_i(-\vec e_i\times\vec P) \bigr).
\end{align}

\subsection{(0,1,1) $A_1$}

We write the frame momentum as $\vec P L/(2\pi) = \vec d_1 + \vec
d_2$, where $\vec d_i=\pm \vec e_j$ for some $j$ and $\vec d_1\perp
\vec d_2$.
\begin{align}
  [BB]_{A_1(0,1,1)}^{0(0,2)} &= [BB]_0(\vec 0), \\
  [BB]_{A_1(0,1,1)}^{0(1,1)} &= [BB]_0(\tfrac{2\pi}{L}\vec d_1), \\
  [BB]_{A_1(0,1,1)}^{1(1,1)} &= \sum_i (\vec d_1\times \vec d_2)_i[BB]_i(\tfrac{2\pi}{L}\vec d_1).
\end{align}
Note that because we only consider flavor symmetric operators, these
are insensitive to the exchange of $\vec d_1$ and $\vec d_2$.

\subsection{(1,1,1) $A_1$}

We write $\vec PL/(2\pi) = \vec d_1 + \vec d_2 + \vec d_3$, where
$\vec d_i=c_i\vec e_i$, $c_i=\pm 1$.
\begin{align}
  [BB]_{A_1(1,1,1)}^{0(0,3)} &= [BB]_0(\vec 0), \\
  [BB]_{A_1(1,1,1)}^{0(1,2)} &= \frac{1}{\sqrt{3}} \sum_i [BB]_0(\tfrac{2\pi}{L}\vec d_i), \\
  [BB]_{A_1(1,1,1)}^{1(1,2)} &= \frac{ c_1c_2c_3 }{\sqrt{6}} \sum_{ijk} \epsilon_{ijk} c_j [BB]_j(\tfrac{2\pi}{L}\vec d_k).
\end{align}

\subsection{(0,0,2) $A_1$}

\begin{equation}
\label{eq:two00}
  [BB]_{A_1(0,0,2)}^{0(1,1)} = [BB]_0(\vec P/2).
\end{equation}

\section{Evaluating correlator matrices using distillation\label{app:distillation}}

As in our previous study~\cite{Francis:2018qch}, we evaluate
correlator matrices involving two-baryon operators using the method
called distillation~\cite{Peardon:2009gh}. In this approach, the
interpolating operators are defined using Laplacian-Heaviside
(LapH)-smeared quark fields. LapH smearing uses the $N_\text{LapH}$
lowest-lying eigenmodes
$\{v_i^{(n,t)}(\vec x) : 1\leq n\leq N_\text{LapH}\}$ of the spatial
gauge-covariant Laplacian (constructed using spatially
stout-smeared~\cite{Morningstar:2003gk} gauge links) on each timeslice
$t$. The smeared quark fields are obtained by projecting onto the
space spanned by these eigenmodes:
\begin{equation}
  \tilde q_i(\vec x,t) \equiv \sum_{n=1}^{N_\text{LapH}} \sum_{j,\vec y} v_i^{(n,t)}(\vec x)
  v_j^{(n,t)*}(\vec y) q_j(\vec y,t).
\end{equation}
LapH smearing is a projector onto a much smaller subspace [in practice
$N_\text{LapH}\ll N_c(L/a)^3$], making it feasible to compute the full
timeslice-to-all quark propagator within this subspace, which is
called the \emph{perambulator}:
\begin{equation}
  \tau_{\alpha\beta}^{n' n}(t,t_0) \equiv \sum_{i,j,\vec x',\vec x}
  v_i^{(n',t)*}(\vec x')
  D^{-1}_{\alpha i,\beta j}(\vec x',t;\vec x,t_0)
  v_j^{(n,t_0)}(\vec x).
\end{equation}
The other key object required for evaluating correlation functions
involving baryons is the \emph{mode triplet},
\begin{equation}
  T_{lnm}(t,\vec p) = \sum_{\vec x}e^{-i\vec p\cdot\vec x}\epsilon^{ijk}
  v_i^{(l,t)}(\vec x) v_j^{(n,t)}(\vec x) v_k^{(m,t)}(\vec x).
\end{equation}

All of our single- and two-baryon correlation functions can be
evaluated by performing tensor contractions of perambulators, mode
triplets, and spin matrices. For a fixed choice of timeslices and
momentum, the perambulator has size $4N_\text{LapH}^2$ and the mode
triplet has size $N_\text{LapH}^3$. (Because of the projector $P_+$ in
our interpolating operators, there are only two independent spin
components.) To keep the smearing width fixed, $N_\text{LapH}$
should be scaled proportional to the spatial lattice volume, and
therefore the scaling of the tensor contraction cost with
$N_\text{LapH}$ should be kept small.

\begin{figure}
  \includegraphics[width=0.35\columnwidth]{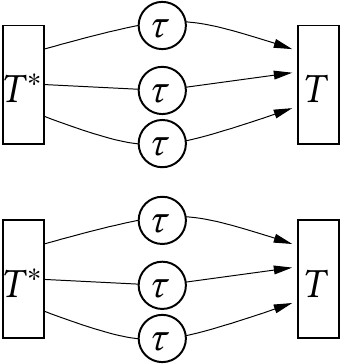}
  \hspace{0.2\columnwidth}
  \includegraphics[width=0.35\columnwidth]{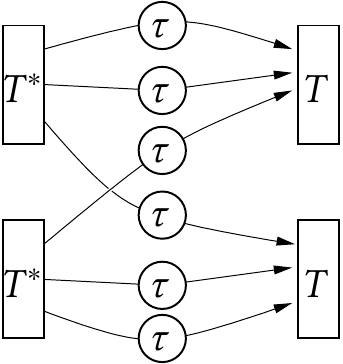}
  \caption{\label{fig:contractions}The two classes of Wick contractions
    for two-baryon correlators, represented as diagrams of tensor contractions
    involving perambulators $\tau$ and mode triplets $T$. Note that
    contractions involving spin indices are not indicated.}
\end{figure}

\begin{figure}
  \includegraphics[width=0.35\columnwidth]{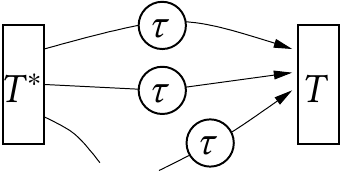}
  \caption{\label{fig:block}Source-sink partially contracted block.}
\end{figure}

The Wick contractions of quark fields yield two topologically distinct
classes of diagrams, shown in Fig.~\ref{fig:contractions}. One
possible strategy would be, in an intermediate step, to construct
two-baryon ``source'' and ``sink'' tensors, where the former is the
outer product of two mode triplets and the latter additionally
includes the six perambulators. This would fully factorize the choice
of source and sink operators in the correlator matrix. However, the
computational cost would scale with $N_\text{LapH}^6$. Instead, we
form partially-contracted source-sink ``blocks''
(Fig.~\ref{fig:block}) at a cost proportional to
$N_\text{LapH}^4$. Computationally, this is the most costly step in the
contractions, and therefore we avoid recomputing blocks that are used
in multiple correlators. The cost of combining two blocks to complete a
two-baryon contraction is proportional to $N_\text{LapH}^2$ and is
relatively inexpensive. A similar strategy for two-baryon correlators
was described recently in Ref.~\cite{Horz:2019rrn}.

In larger volumes, the $N_\text{LapH}^4$ cost scaling will eventually
become prohibitively expensive. One possible solution is to use
stochastic distillation~\cite{Morningstar:2011ka, Horz:2020zvv}, which would
replace $N_\text{LapH}$ in the cost scaling with the (much smaller)
size of the dilution space.

\subsection{Choosing $N_\text{LapH}$\label{sub_sub_sec:n_laph}}

\begin{figure*}
  \includegraphics[width=0.48\textwidth]{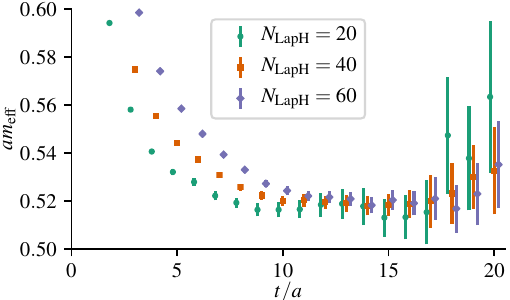}
  \hfill
  \includegraphics[width=0.48\textwidth]{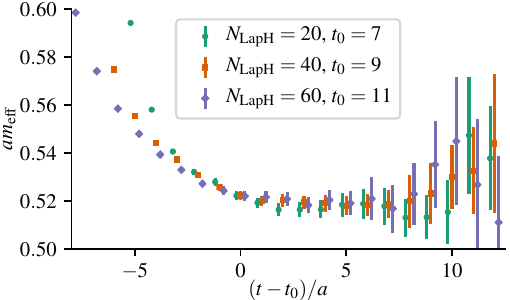}
  \caption{\label{fig:N_laph_dependence}\textbf{Left}: Effective energies 
    for an octet baryon correlator on U103 with $N_\text{LapH} = 20,40,60$.
    \textbf{Right}: The effective energies shifted such that their plateaux
    start at $t-t_0=0$.}
\end{figure*}

Due to the rise in inversion and contraction costs as $N_\text{LapH}$ is increased,
it is computationally advantageous to use as few LapH eigenvectors as possible. However, making
$N_\text{LapH}$ too small will increase the statistical uncertainty.
Hence, for comparison, we computed an octet-baryon correlation function using three values of $N_\text{LapH}$
on a subset of ensemble U103. The effective energies are shown in
the left panel of Fig.~\ref{fig:N_laph_dependence}.
It is clearly seen that the error on the effective energy
increases as the number of LapH eigenvectors is reduced.
At the same time, retaining fewer LapH eigenmodes has 
resulted in less contamination from the excited states;
therefore, a more fair comparison between the three is one in which the onset of the
plateau for each effective energy has been shifted to the same point. This
is shown in the right panel of Fig.~\ref{fig:N_laph_dependence}, indicating
that $N_\text{LapH} = 20$ is an acceptable choice. For the other
$N_f=3$ ensembles, $N_\text{LapH}$ is scaled with the physical
three-volume to ensure that the smearing radius remain roughly
constant. For the $N_f=2$ ensembles, we have a single volume and we
choose to use a slightly larger $N_\text{LapH}$, corresponding to a
smaller smearing radius.

\section{Analysis of correlation functions\label{app:corr}}

\begin{figure}
  \centering
  \includegraphics[width=\columnwidth]{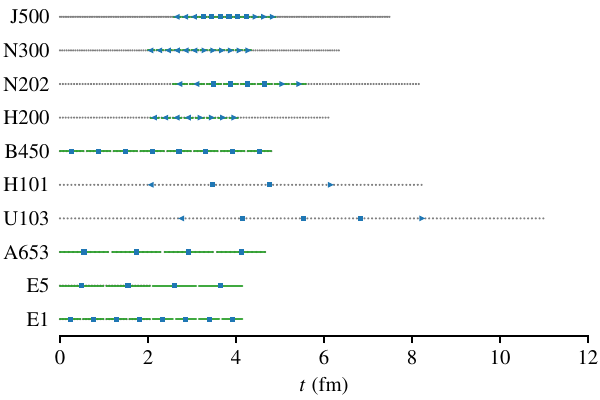}
  \caption{Location of source times on all ensembles. Triangles
    indicate sources used for only forward-propgating or
    backward-propagating states, and squares indicate sources used for
    both. When present, green line segments indicate the range over
    which sources were randomly shifted on each gauge configuration.}
  \label{fig:sources}
\end{figure}

The correlation functions computed are of the form
\begin{equation}
  \label{eq:corr_mat}
  C_{ij} (t) \equiv \frac{1}{N_t} \sum_{\left\{t_{0}\right\}} \langle \mathcal{O}_i (t + t_0) \mathcal{O}_j^\dagger (t_0) \rangle ,
\end{equation}
where $\left\{\mathcal{O}_i\right\}$ denotes a set of interpolating operators that all transform irreducibly in the same way,
and $\left\{t_{0}\right\}$ is the set of $N_t$ sources shown in
Fig.~\ref{fig:sources} for all ensembles.
The sources and time separations that we include assume
$t \ll T$ for periodic boundary conditions, and both $0 \ll t_0$ and $t+t_0 \ll T$ for open boundary conditions,
such that the effects of the finite temporal extent may be ignored.
Under these assumptions, the spectral decomposition of the correlators is given by
\begin{equation}
  C_{ij} (t) = \sum_{n=0}^{\infty} \langle \Omega | \mathcal{O}_i | n \rangle \langle \Omega | \mathcal{O}_j | n \rangle^\ast e^{-E_n t} ,
\end{equation}
where $| \Omega \rangle$ is the vaccuum state, $| n \rangle$ are the eigenstates of the system, and $E_n$ are the eigenenergies.

\subsection{Octet-baryon mass}

\begin{figure}
  \includegraphics[width=\columnwidth]{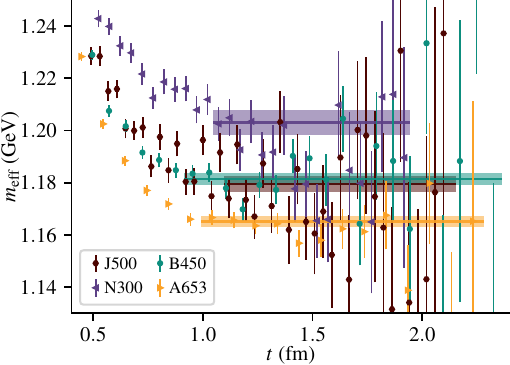}
  \caption{Effective energy for the octet baryon with total momentum zero on
    four ensembles with similar volumes. The bands show the value of the
    octet-baryon mass obtained from single-exponential fits to a single-octet-baryon
    correlator projected to zero momentum and also indicate the range
    of $t$ used for the fit.}
  \label{fig:baryon_eff_energy}
\end{figure}

In order to calculate $p^2$, which is needed for the phase-shift analysis, we must obtain an estimate
for the octet-baryon mass. To this end, we perform
single-exponential fits to correlators constructed from a single-octet-baryon operator projected to zero momentum.
We show the resulting fits and effective energies on four ensembles with similar volumes in Fig.~\ref{fig:baryon_eff_energy}.

\subsection{Generalized eigenvalue problem}
\label{sec:gevp}

\begin{figure*}
  \includegraphics[width=0.48\textwidth]{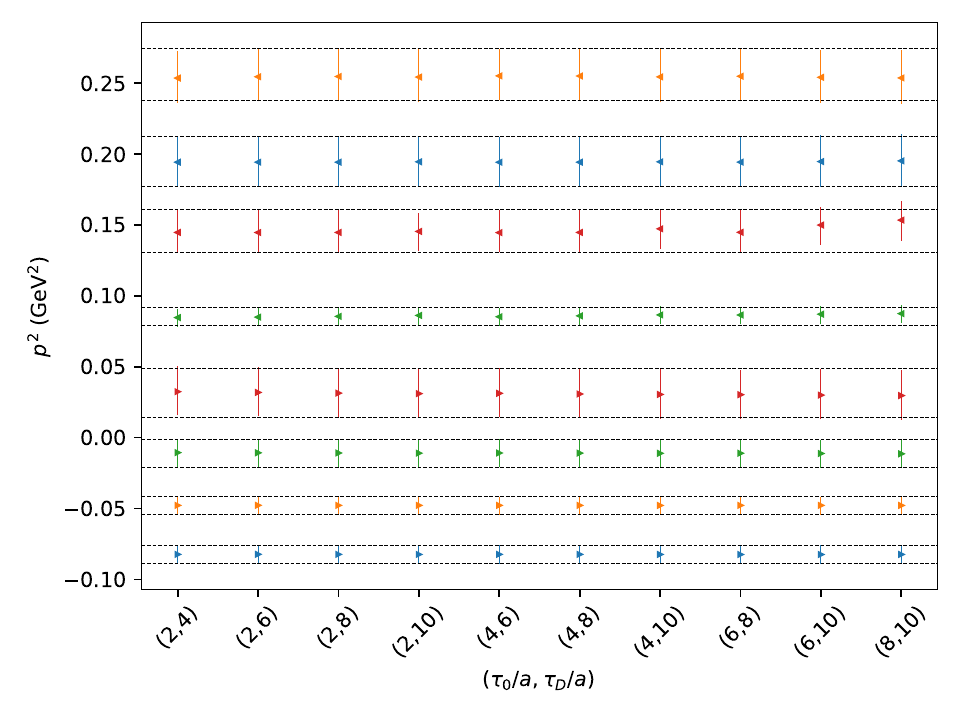}
  \hfill
  \includegraphics[width=0.48\textwidth]{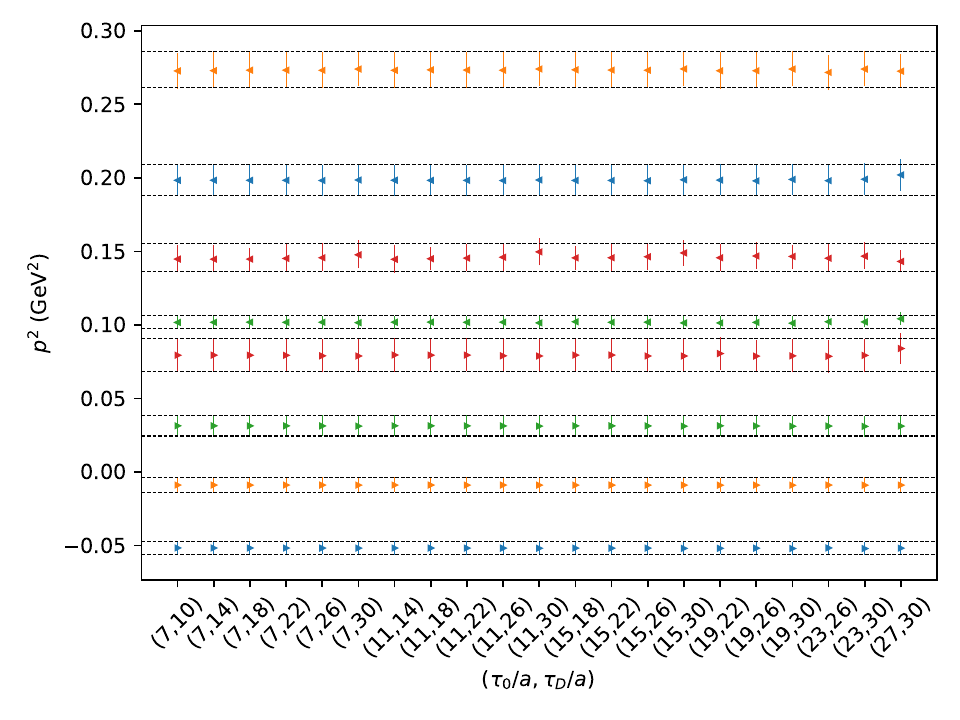}
  \caption{\label{fig:gevp}The center-of-mass scattering momentum $p^2$ versus the GEVP parameters $(\tau_0, \tau_D)$ for
    A653 (left) and J500 (right). The momentum frames include $\vec PL/(2\pi)=(0,0,0)$ (blue), $(0,0,1)$ (orange), $(0,1,1)$ (green), and $(1,1,1)$ (red).
    The ground states are denoted with right-facing triangles and the excited states with left-facing triangles.
  The second excited state in the rest frame and the spin-one states are not included.
  The dashed black lines show the upper and lower errors for each energy from the chosen values of $(\tau_0/a, \tau_D/a)$, which are $(4,8)$ for A653 and $(11,22)$ for J500.}
\end{figure*}

For all momentum frames that include more than one two-baryon operator, we use the variational approach described
in Refs.~\cite{Luscher:1990ck,Blossier:2009kd}, in which a generalized eigenvalue problem (GEVP) is solved
from the matrix of correlation functions in Eq.~\eqref{eq:corr_mat}:
\begin{equation}
  C(t) \upsilon_n (t, \tau_0) = \lambda_n (t, \tau_0) C(\tau_0) \upsilon_n (t, \tau_0).
\end{equation}
Provided that $\tau_0$ satisfies $\tau_0 \ge t/2$,
the asymptotic behavior of the generalized eigenvalues is given by \cite{Blossier:2009kd}
\begin{equation}
  \lambda_n (t) = |A_n|^2 e^{-E_n t}\left[ 1 + O(e^{-(E_N - E_n)t}) \right],
\end{equation}
where $N$ is the size of the correlator matrix, and
the argument $\tau_0$ has been dropped. By
contrast, the leading corrections to the eigenvalues of $C(t)$ only fall
off as $e^{-t \Delta_n}$, where $\Delta_n \equiv \min_{m \neq n} | E_n - E_m |$.
Thus, by solving the GEVP rather than the simple eigenvalue problem
for $C(t)$, one benefits from a stronger suppression of the
contamination from higher excitations.

To simplify the analysis, we turn the GEVP into a normal eigenvalue problem, resulting in
the following matrix to be diagonalized
\begin{equation}
  \hat{C}(t) \equiv C^{-1/2} (\tau_0) C(t) C^{-1/2} (\tau_0) ,
\end{equation}
and only solve for the eigenvectors and eigenvalues at a single time separation $\tau_D > \tau_0$.
The resulting eigenvectors can be used to rotate $\hat{C}(t)$ for all other time separations
\begin{equation}
  \tilde{C}(t) \equiv V^\dagger(\tau_D) C^{-1/2} (\tau_0) C(t) C^{-1/2} (\tau_0) V(\tau_D) ,
\end{equation}
where the columns of $V(\tau_D)$ contain the orthonormal eigenvectors of $\hat{C}(\tau_D)$.
Then the diagonal elements of $\tilde{C}(t)$ approximate the generalized eigenvalues $\lambda_n (t)$.
It can be seen in Fig.~\ref{fig:gevp} that the scattering momenta derived from the spectrum show very little dependence on the chosen GEVP parameters $\tau_0$ and $\tau_D$.
The rotated correlators are inspected by eye to ensure they remain statistically diagonal
for all time separations.

Finally, extraction of the leading exponential terms for the diagonal elements of $\tilde{C}(t)$
gives the lowest $N$ levels that overlap with the states created by the operators used
in the correlation matrix, and the overlaps themselves are given by
\begin{equation}
  \label{eq:overlap_factors}
  Z_j^{(n)} \equiv \bra{0} \mathcal{O}_j \ket{n} \approx C_{j k}^{1/2} (\tau_0) V_{k n} (\tau_D) A_n .
\end{equation}
These overlaps are used to identify states as being predominantly spin-zero or spin-one.

\subsection{Ratio fits}

In a final step before fitting the correlators,
we form a ratio of each diagonal element of the rotated correlator matrix
to the product of two single-baryon correlators,
\begin{equation}
  R_n (t) \equiv \frac{\tilde{C}_{nn}(t)}{C_{\Lambda}^{\vec{p}_1} (t) C_{\Lambda}^{\vec{p}_2} (t)} .
\end{equation}
The momenta $\vec p_{1,2}$ are chosen to correspond to the constituent
momenta of the individual baryons appearing in the operator that has
dominant overlap with state $n$.
The advantage of forming this ratio is the possibility for partial cancellation of correlations and residual contributions from excited states.
One drawback, however, is the loss of the monotonic behavior of the effective energy, making an identification of the plateau less reliable.
To avoid this issue, in most cases we fix the lower end of the fit range, $t_{\rm min}$, on each ensemble to the first time separation in the plateau region of the single-baryon correlators,
which were observed to take longer to reach their asymptotic behavior than the two-baryon correlators.

However, as mentioned in the main text, in some cases this choice of
$t_{\rm min}$ corresponds to a poor signal quality in $R_n(t)$ and we
instead chose a slightly lower $t_{\rm min}$.  For all of these levels
that are also used in the phase shift analysis, the decrease of
$t_{\rm min}$ below the start of the single-baryon plateau was by less
than 0.12~fm, except on ensemble N300 where the decrease was by
0.25~fm. These choices still lie in the plateau region of $R_n(t)$.

\begin{figure*}
  \includegraphics{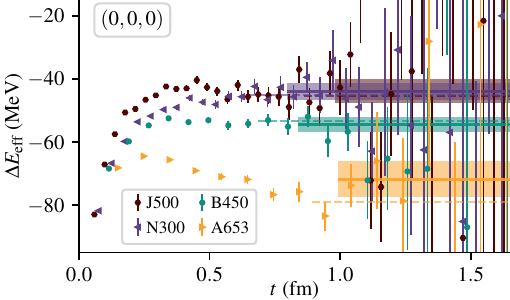}
  \includegraphics{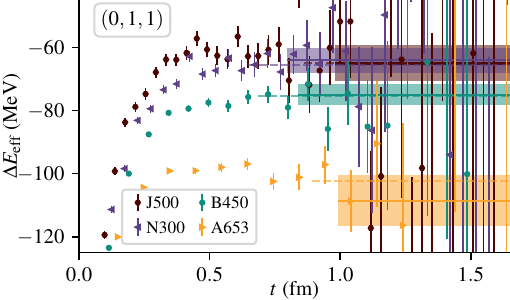}
  \caption{Effective energy differences for the ground state in frames
    $(0,0,0)$ (left) and $(0,1,1)$ (right) on four ensembles with
    similar volumes. See the caption of Fig.~\reffigeffenergy.}
  \label{fig:eff_energy_P0_P2}
\end{figure*}

To estimate the systematic error corresponding to the chosen fit
range, we extracted an alternative spectrum, based on a second value
of $t_{\rm min}$ that is below our preferred value by somewhere
between 0.0865--0.173~fm, and propagated it through to the subsequent
analysis.
Finally, the upper end of the fit range, $t_{\rm max}$, is chosen for each correlator ratio to be one time separation smaller than the first time separation in which $|R_n(t)| < 3 \; \text{error} (R_n(t))$.
Effective energy differences for two additional ground-state levels are shown in Fig.~\ref{fig:eff_energy_P0_P2}.

By fitting the ratio $R_n (t)$, one obtains the shift
$\Delta E$ of the $n$th interacting energy eigenstate relative to the
corresponding noninteracting level. The interacting energy $E$ is
then reconstructed from $\Delta E$ by adding the noninteracting energy level,
$\sqrt{p_1^2+m_B^2}+\sqrt{p_2^2+m_B^2}$,
determined from the continuum dispersion relation using the single-octet-baryon energy at rest.

\section{Combined fits\label{app:combined}}

We begin by describing the selection of levels that are included in
the combined fits. In general, an energy level is excluded for one of
four reasons:
\begin{enumerate}
\item Energy levels with dominant coupling to spin-one interpolating
  operators are excluded. The corresponding partial waves such as
  ${}^3P_1$ factorize in the quantization condition.
\item The spin-zero excited state in frames $(0,1,1)$ and $(1,1,1)$
  cannot be described using the simplest form of the quantization
  condition. Assuming the phase shift does not pass through zero,
  Eq.~\refeqquantization\ has a solution between the lowest pair of
  noninteracting levels, whereas the data lie below this
  range. Examining these levels in the nonrelativistic limit, one sees
  that they belong to the same degenerate shell of states, which can
  contain just one ${}^1S_0$ level. Therefore, higher partial waves
  are relevant. These levels can be described if ${}^1D_2$ is included
  in the quantization condition, which we leave to future
  work~\cite{Green_Lat21}; they are excluded from all analyses here.
\item Energy levels with too-high $p^2$ are susceptible to the influence
  of the three-particle inelastic threshold, which is described by
  neither our fit ansatz nor the quantization condition. Therefore we
  exclude the second excited state in frame $(0,0,0)$ on all ensembles
  except for the two largest volumes, H101 and N202.
\item Energy levels with too-low $p^2$ are susceptible to the
  influence of the $t$-channel cut (arising from the exchange of a
  pseudoscalar meson), which is described by neither our fit ansatz
  nor the quantization condition. (We note that the method recently
  proposed in Ref.~\cite{Meng:2021uhz} might be applicable.)
  However, on our coarser ensembles
  the bound-state pole also lies close to the $t$-channel cut. The
  ground state in frame $(0,0,1)$ is essential for constraining the
  pole position, and therefore we always include it, even though on
  our coarsest lattice spacing this level lies below the cut.

  On the other hand, for almost all ensembles the ground states in
  frames $(0,0,0)$ and $(0,0,2)$ lie below the cut and we exclude
  these levels. The exception is the largest volume, N202. However,
  these two levels still lie well below the bound-state pole and are
  very close to the cut; furthermore, we obtain significantly worse
  fit quality when either of these two levels is included. (For
  instance, the single-ensemble fit to the low-$p^2$ region of N202
  has $\chi^2/\text{dof}=1.7/1$. Including the ground state in the
  rest frame increases this to 12.0/2. For the fit to the full-$p^2$
  range, $\chi^2/\text{dof}$ increases from 4.1/3 to 8.2/4 when
  including this level.)  Therefore, we also exclude these two levels
  on N202.
\end{enumerate}
Our final choice of levels for the full $p^2$ range is the following:
one or two excited-state levels in frame $(0,0,0)$, both the ground
and excited spin-zero levels in frame $(0,0,1)$, and the ground state
in frames $(0,1,1)$ and $(1,1,1)$. For the near-threshold region, we
take the ground state in frame $(0,0,1)$ and possibly the ground state
in frames $(0,1,1)$ and $(1,1,1)$.

The fits are performed by minimizing
\begin{equation}
  \chi^2 \equiv \sum_{i,j} ( p^2_i - p^2_{\text{q.c.},i} )
   \Sigma^{-1}_{ij} ( p^2_j - p^2_{\text{q.c.},j} )
\end{equation}
with respect to the model parameters, where $i$ indexes all of the
levels among all ensembles included in the fit and
$p^2_{\text{q.c.},i}$ is obtained by solving
Eq.~\refeqquantization\ given the model for $p\cot\delta(p)$. Here
$\Sigma=\Sigma_\text{stat}+\Sigma_\text{syst}$ is an estimate of the
covariance matrix. Bootstrap resampling is used to obtain
$\Sigma_{\text{stat},ij}$, which is set to zero when $i$ and $j$
correspond to levels from different ensembles. The alternative
spectrum fit range is used to estimate a correlated systematic
uncertainty: we set $\Sigma_{\text{syst},ij}=(\delta p^2)_i (\delta
p^2)_j$, where $\delta p^2$ is the the difference between $p^2$
obtained using the preferred and alternative spectra.

The statistical uncertainty of our fit results is estimated using
bootstrap. When fitting to the near-threshold region, for a small
number of bootstrap resamples (up to 4 out of 1000) the minimum of
$\chi^2$ is not a point where its gradient vanishes, but instead lies
at a discontinuity. In these rare cases, there exists a level
(typically the lowest-lying level in the smallest volume) where the
left-hand and right-hand sides of Eq.~\refeqquantization\ are
tangent and a small adjustment of the model parameters causes the
solution to disappear. Although this represents a breakdown of the
quantization condition and/or unphysical model parameters, we still
keep these solutions in our statistical analysis as their effect is
negligible. In addition to the bootstrap resamples, we also perform an
additional fit using the alternative spectrum and take the difference
in fit results as an estimate of systematic uncertainty.

A similar problem occurs for the bootstrap estimate of the
uncertainty of the interacting spectrum in the continuum obtained
using Eq.~\refeqquantization\ and shown in Fig.~\reffigspectra. When
$L$ is small, for some of the samples the ground state solution in
frames $(0,0,0)$ and $(0,0,2)$ disappears. Because of this, we do not
show an error band for these cases, which correspond roughly to
energies below the $t$-channel cut.

\begin{figure}
  \includegraphics{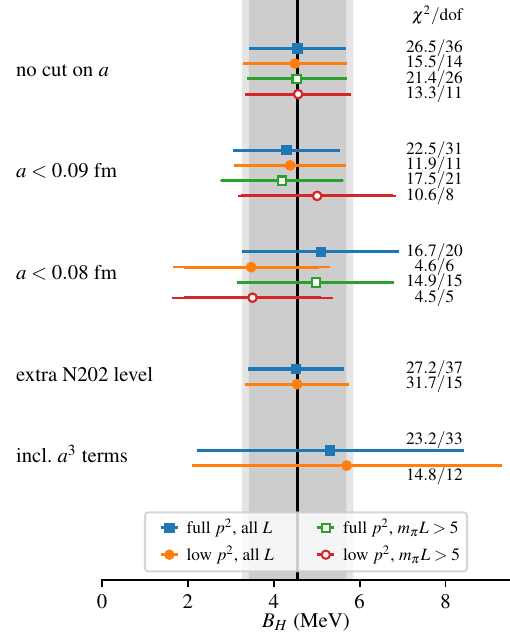}
  \caption{Binding energy, $\chi^2$, and number of degrees of freedom
    for various combined fits. The thin outer error bar includes the
    (usually negligible) estimate of systematic uncertainty based on
    the alternative spectrum, added in quadrature. Fits to the full
    $p^2$ range are indicated by squares and those to the
    near-threshold region by circles. Open symbols correspond to fits
    that exclude the two ensembles with $L\approx 2.1$~fm. The first
    three groupings exclude the zero, one, and two coarsest lattice
    spacings. The last two groupings are variations on the fits that
    include all ensembles, either adding the rest-frame ground state
    from N202 or parametrizing the fit coefficients as
    $c_i = c_{i0} + c_{i1} a^2 + c_{i2} a^3$. The vertical line with
    inner and outer error bands show our final estimate,
    Eq.~\refeqbindEfinal.}
  \label{fig:extrap_summary}
\end{figure}

To estimate additional systematic uncertainty due to the continuum
extrapolation and residual finite-volume effects, we apply various
cuts to the selection of ensembles. In addition, to probe the ansatz
for $p\cot\delta$, we use both a quadratic polynomial in $p^2$ with
the full $p^2$ range and a linear polynomial with the near-threshold
region. These fits are summarized in the first three groupings of
Fig.~\ref{fig:extrap_summary}. When fitting to all six lattice
spacings, the resulting binding energy is very stable with respect to
the inclusion of the small volumes and the choice of $p^2$ range. As
the coarser lattice spacings are excluded, the variations increase,
with the choice of $p^2$ range becoming more important than the cut on
$L$. Our preferred fit, which provides our central value and
statistical uncertainty, is the one that includes the most data, i.e.\
the first in the figure. We estimate the systematic uncertainty as the
root-mean-square difference from the preferred fit of the central
values of the seven other fits in the first and third groupings.

Figure~\ref{fig:extrap_summary} also shows two additional
variations. Including the ground state in the rest frame from N202 has
a negligible impact on the binding energy but can substantially
increase $\chi^2$. Including $a^3$ terms in the dependence on the
lattice spacing significantly increases the uncertainty, without
improving the fit quality.

\begin{figure}
  \includegraphics{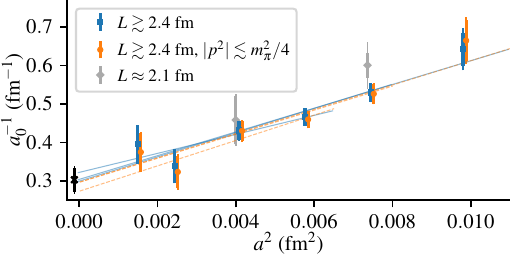}
  \includegraphics{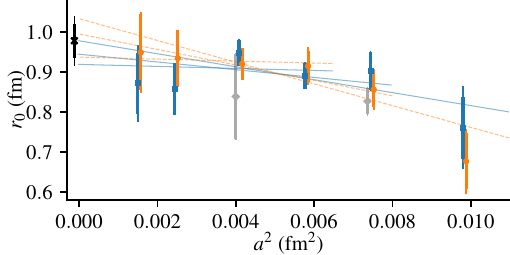}
  \caption{Inverse scattering length versus squared lattice spacing
    (top) and effective range versus squared lattice spacing
    (bottom). See the caption of Fig.~\reffigbindEvsa.}
  \label{fig:a0r0_vs_a}
\end{figure}

The curves showing the dependence of $B_H$ on $a^2$ in
Fig.~\reffigbindEvsa\ are based on the fits whose results are
shown as filled blue squares and orange circles in the first three
groupings of Fig.~\ref{fig:extrap_summary}. The same is shown for
$a_0^{-1}$ and $r_0$ in Fig.~\ref{fig:a0r0_vs_a}. The inverse
scattering length shows a strong dependence on the lattice spacing
(varying by a factor of two) and is fairly insensitive to the choice
of $p^2$ range. The effective range has a weaker dependence on the
lattice spacing but shows larger variation with the choice of $p^2$
range; this contributes to its relatively larger systematic
uncertainty.

\section{Two-flavor ensembles\label{app:Nf2}}

In addition to our main analysis of $N_f=3$ lattice ensembles, we have
generated new data for two $N_f=2$ ensembles (i.e.\ with dynamical $u$
and $d$ quarks and a quenched $s$ quark) used in our previous study of
the $H$ dibaryon~\cite{Francis:2018qch} and listed in the lower part
of Table~\ref{tab:ensembles}. Based on the analysis in
Ref.~\cite{Draper:2021clv}, we expect that the quenched $s$ quark is
not an obstacle to using finite-volume quantization conditions.
On both ensembles we elected to set the
strange quark mass equal to that of the light quarks; this means that
both ensembles have SU(3) flavor symmetry in the valence sector. For
ensemble E5 with a pion mass of 437~MeV, this is a change from
Ref.~\cite{Francis:2018qch} where we tuned the strange quark mass to
be near its physical value; as a result, the main difference between
E5 and the $N_f=3$ ensembles is that the strange quark is quenched.

\begin{figure*}
  \centering
  \includegraphics[width=0.48\textwidth]{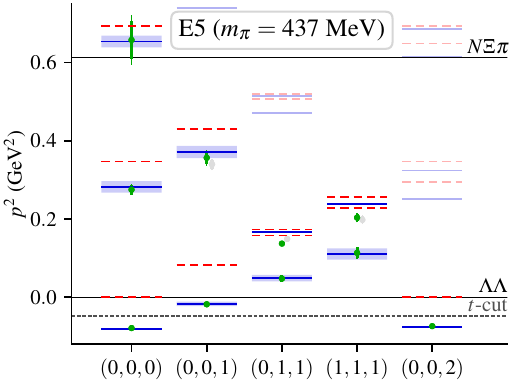}
  \hfill
  \includegraphics[width=0.48\textwidth]{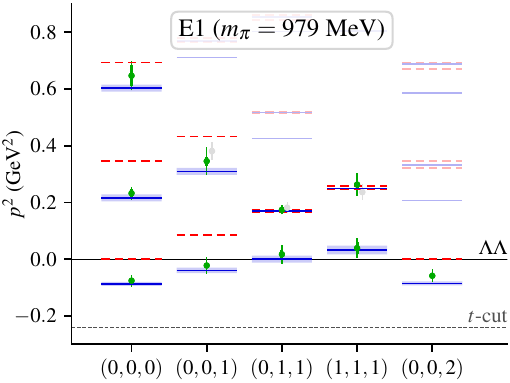}
  \caption{Two-baryon spectrum in five different reference frames on
    ensembles E5 (left) and E1 (right). Green points are the spin-zero
    levels and gray points are the spin-one levels. Horizontal lines
    indicate two- and three-particle thresholds and the $t$-channel cut.
    Horizontal line segments show finite-volume energies in the
    noninteracting case (dashed red) and from the fit to the wider $p^2$
    range (solid blue).}
  \label{fig:spectrum_E5_E1}
\end{figure*}

Our analysis on the $N_f=2$ ensembles is the same as what was done in
the $N_f=3$ case, except that we cannot study the continuum limit. The
finite-volume spectra obtained from ensemble E5 are shown in
the left panel of Fig.~\ref{fig:spectrum_E5_E1}; they have the same
qualitative features as observed for the $N_f=3$ ensembles. Performing
fits of the phase shift, we obtain a binding energy
\begin{equation}
\label{eq:E5}
  B_H = 12.0 \pm 2.7 \pm 0.5 \text{ MeV}\quad \text{(E5)},
\end{equation}
which is consistent with the binding energies in the $N_f=3$ case at
similar nonzero lattice spacing.

The right panel of Fig.~\ref{fig:spectrum_E5_E1} shows the
finite-volume spectra for ensemble E1. As the pion mass is much
larger, the $t$-channel cut and three-particle threshold are further
away from the threshold and all of the obtained levels lie in the
region where the two-particle quantization condition is applicable.
On this ensemble, the uncertainty of both the spectrum and the fitted
quantities are dominated by systematics. The phase shift fits yield
\begin{equation}
\label{eq:E1}
  B_H = 17.3 \pm 4.0 \pm 5.4 \text{ MeV}\quad \text{(E1)},
\end{equation}
which is consistent with the value $19\pm 10$~MeV reported in our
previous work~\cite{Francis:2018qch} but has a smaller error.

In Fig.~\reffigcomparison\ we compare our results for the binding
energy $B_H$ for $N_f=3$ and $N_f=2$ with the estimates from HAL~QCD
\cite{Inoue:2010es,Inoue:2011ai} and NPLQCD
\cite{Beane:2010hg,Beane:2011zpa,Beane:2012vq}. Our $N_f=2$
calculations at nonzero lattice spacing show a dependence on the pion
mass compatible with that observed by HAL~QCD, although they lack the
precision necessary to make an unambiguous statement. Moreover, this
plot underscores our observation that discretization effects in this
quantity are sizeable.

\section{Spectrum data}\label{sec:data}

The spectra used in this work are available in HDF5 format~\cite{hdf5}
in the file \texttt{levels.h5}. Each dataset contains the bootstrap
samples for one or more energy levels in lattice units, with the
ensemble, frame, and type of energy level specified by the dataset's
key. For example, the following correspond to ensemble N300 and
frame $\vec PL/(2\pi)=(0,1,1)$:
\begin{verbatim}
/N300/P011/octet_baryon  Dataset {1001},
/N300/P011/spin_one      Dataset {1001, 2, 1},
/N300/P011/spin_zero     Dataset {1001, 2, 2}.
\end{verbatim}
The first entry of the first dimension contains the average over the
ensemble and the next 1000 are the bootstrap samples. For the
two-baryon spectrum, the second dimension indexes the preferred and
alternative values in the first and second entries, and the third
dimension indexes the energy levels in ascending order.
As should be evident from the names, the first of these three datasets contains
the octet baryon energy with momentum $\vec P$, the second contains
the two-baryon level identified as spin one, and the last contains
both two-baryon-levels identified as spin zero.

\end{document}